\documentclass[12pt]{iopart}

\begin{document}
\title[
%i Non-Conservation of the
Evolution of the
Moment of Inertia 
of Figure-Eight Choreography
]{
%Saari-Chenciner's Theorem
%for Three Body Figure-Eight Choreography\\
%or\\
%Proof of 
%i Non-Conservation of the
Evolution of the
Moment of Inertia of 
Three-Body Figure-Eight Choreography
}
\author{Toshiaki Fujiwara\dag, Hiroshi Fukuda\ddag\ and Hiroshi Ozaki\P}
\address{\dag\ Faculty of General Studies, Kitasato University, 
Kitasato 1-15-1, Sagamihara, Kanagawa 228-8555, Japan}%
\address{\ddag\ School of Administration and Informatics,
University of Shizuoka, 
52-1 Yada, Shizuoka 422-8526, Japan}
\address{\P\ Department of Physics, Tokai University,
1117 Kitakaname, Hiratsuka, Kanagawa 259-1292, Japan}

\eads{
	\mailto{\dag\ fujiwara@clas.kitasato-u.ac.jp},
	\mailto{\ddag\ fukuda@u-shizuoka-ken.ac.jp},
	\mailto{\P\ ozaki@keyaki.cc.u-tokai.ac.jp}
}

%abstract%%%%%%%%%%%%%%%%%%%%%%%
\begin{abstract}
We investigate
three-body motion 
in three dimensions
under  
%homogeneous potential
the interaction potential 
%energy 
proportional to 
$r^\alpha$
($\alpha \neq 0$) or
$\log r$,
where $r$ represents the mutual distance
between bodies,
with the following conditions:
(I) the moment of inertia is non-zero constant,
(II) the angular momentum is zero,
and
(III) one body is on the centre of mass at an instant.

We prove that
the motion which satisfies conditions (I)--(III)
with equal masses
for
$\alpha \neq -2, 2, 4$
%is the motion with zero moment of inertia
%must have zero moment of inertia
%with respect to the centre of mass,
%i.e., 
%triple collision forever.
%
is impossible.
And
motions which satisfy the same conditions
for  $\alpha = 2, 4$
are solved explicitly. 
Shapes of these orbits are not figure-eight
and these motions have collision.
%
%
%Therefore non-conservation of
%the moment of inertia 
%for figure-eight choreography 
%for $\alpha \neq -2$
%is proved.
%i.a
Therefore
the moment of inertia for figure-eight choreography
for $\alpha \neq -2$ is proved
to be inconstant along the orbit.

%since
%it satisfies the conditions (II) and (III)
%and has no collision.

%
We also prove that
the motion which satisfies conditions (I)--(III)
with general masses
under the Newtonian potential 
$\alpha = -1$
is impossible.
%is only the motion with
%triple collision forever.
%
\end{abstract}


%\maketitle
%
% body
%
%Introduction%%%%%%%%%%%%%%%%%%%%%%%%%
\section{Introduction}
In 1970's,
Saari formulated a conjecture
\cite{saari, xia},
which is now called ``Saari's Conjecture'':
{\it
In the n-body problem
under the Newtonian gravity,
if the moment of inertia is constant
then the motion must be 
a relative equilibrium.
}
Recently,
three-body choreography,
equal mass
three-body periodic motion on a planer closed curve
on which each body chase each other,
was found by
Moore \cite{moore}, 
Chenciner, Montgomery \cite{chenAndMont}
and Sim\'{o} \cite{simo1, simo2}.
This motion is now called 
``three-body figure-eight choreography''.
%%
%%One of the interesting properties
%%of this motion is that
%%the moment of inertia is 
%%almost constant,
%%which is consistent to
%%the Saari's conjecture.
Sim\'{o} noticed that the moment of 
inertia was not constant on  figure-eight solution for the
Newtonian potential, despite the relative variation along the orbit is
small \cite{simo3}.
%Non-conservation
Inconstancy of the moment of inertia of figure-eight solution
is consistent to the Saari's Conjecture.

On the other hand,
it is well known that
in the n-body problem
%under the homogeneous potential $r^{-2}$
%under the attractive potential $-r^{-2}$
under the attractive potential proportional to $r^{-2}$,
where $r$ is the mutual distance
between bodies,
the moment of inertia $I$
for any periodic motion
must be constant.
This is because the second derivative of the moment of inertia
with respect to time
under this potential
yields the Lagrange-Jacobi identity
%Lagrange-Jacobi identity
$d^{2}I/dt^2=2E$,
where $E$ represents
the total energy.
Integrating this equation,
we get
$I = Et^2+c_{1}t+c_{2}$
with integration constant $c_{1}$ and $c_{2}$.
For any periodic motion
under this potential,
therefore,
the total energy must be zero
and 
the moment of inertia must be constant.
%
%
%Three-body figure-eight choreography
%exists under this potential
%\cite{moore}\cite{simo1}
%%This choreography 
%and
%must have constant moment of inertia.
%

Numerical evidence of existence of 
three-body figure-eight choreography
is known
under the attractive interaction potential 
proportional to $r^{\alpha}$
($\alpha \ne 0$) 
with $\alpha < 2$ or $\log r$ \cite{moore}.
Then, Chenciner formulates a problem
\cite{chenq13}:
{\it
Show that the moment of inertia
of figure-eight choreography
stays constant
only when $\alpha=-2$.
}
%
%ii.5 In this paper,
%ii.5 we solved  this problem.
%ii.5 Therefore we would like to
%ii.5 call this theorem ``Saari-Chenciner's Theorem''.
We call this problem Saari-Chenciner's problem.
In this paper,
we solved  this Saari-Chenciner's problem.

Actually,
we investigated
the three-body motion 
in three dimensions
under the  
%homogeneous potential $r^{\alpha}$ or $\log r$
attractive interaction potential 
proportional to
$r^{\alpha}$ ($\alpha \ne 0$) or $\log r$
with the following conditions:
(I) the moment of inertia is non-zero constant,
(II) the angular momentum is zero,
and
(III) one body is on the centre of mass at an instant.
We proved 
Theorem 1:
{\it Motion
which satisfies the conditions (I)--(III)
with equal masses
under the potential 
$\alpha \neq -2,2,4$
%must have
%zero moment of inertia
%with respect to the centre of mass.%
is impossible.
}
%
%That is, the only possible motion is that
%three bodies stay at the centre of mass
%forever, 
%in other words,
%triple collision forever.
%We call this motion 
%the trivial motion.
%

%And 
We solved explicitly 
%non-trivial 
motions
which satisfy the conditions (I)--(III)
with equal masses
under the potential $\alpha = 2$ or $4$,
and show that these motions do not have
figure-eight shape
and have collision.
Since three-body figure-eight choreography
satisfies the conditions (II)--(III)
with equal masses
and have no collision,
%ii.5 the ``Saari-Chenciner's Theorem''
the Saari-Chenciner's problem
is 
%ii.5 proved.
solved.
%
%Situations are the same 
%for Sim\'{o}'s H3 orbit
%\cite{simo2}.
%Therefore the H3 orbit too
%does not have constant
%moment of inertia.

We also proved 
Theorem 2:
{\it Motion
which satisfies the conditions (I)--(III)
with general masses
under the Newtonian potential 
$\alpha = -1$
%is only the trivial motion.%
is impossible.
}

%Construction
Construction of this paper is as follows.
In section \ref{section:prep}
we clarify the consequences
of the conditions (I)--(III)
for general masses and 
general $\alpha$.
Prescription 
% or Method ?
of our proof of
the Theorem 1 and 2 are given
in this section.
In section \ref{section:equalmass}
we treat the case of equal masses.
In section \ref{section:maincase}
a proof of the Theorem 1 is given.
In section \ref{section:alpha24}
we give 
%non-trivial 
motions explicitly 
which satisfy the conditions (I)--(III)
with equal masses
under the potential $\alpha = 2$ or $4$,
and show that these solutions do not have
figure-eight shape,
and have collision.
In section \ref{sec:nonequalmass}
we treat the case with general masses
under the Newtonian potential $\alpha=-1$,
and give a proof of the Theorem 2.
Summary and discussions are given
in section \ref{section:summary}.
Some algebraic details
for the section \ref{section:maincase}
are shown in \ref{app:comroot}.
%

%new section%%%%%%%%%%%%%%%%%%%%%%
\section{Consequences of the conditions (I)--(III)}
\label{section:prep}
In this section
we clarify the consequences of the conditions (I)--(III)
with general masses and general $\alpha$,
and give prescription 
of our proof of
the Theorem 1 and 2.

Let us consider the three-body problem
in three dimensional space.
Let $m_{i}$ be masses of bodies $i=1,2,3$, 
and let
$\mathbf{r}_{i}(t)$ and $\mathbf{v}_{i}(t)$ 
be position and velocity vectors of them at time $t$,
respectively.
The moment of inertia with respect to the origin $I$, 
the kinetic energy $K$ and
the angular momentum $L$
are defined as follows,
\begin{eqnarray}
I =  \frac{1}{2}\sum_{i} m_{i} \mathbf{r}_{i}^2,
\label{eq:I}
\\
K = \frac{1}{2}\sum_{i} m_{i} \mathbf{v}_{i}^2,
\label{eq:K}
\\
L  = \sum_{i} m_{i}\mathbf{r}_{i} \times \mathbf{v}_{i}.
\label{eq:L}
\end{eqnarray}
To treat the power-law %potential 
and logarithmic potentials
uniformly,
we use the following expression for the potential energy,
\begin{eqnarray}
V_{\alpha}
&=&\alpha^{-1}\sum_{i>j} m_{i}m_{j}r_{ij}^{\alpha} \mbox{  for $\alpha \neq 0$},
	\nonumber\\
&=& \sum_{i>j} m_{i}m_{j}\log r_{ij} \mbox{  for $\alpha =0$},
\label{eq:V}
\end{eqnarray}
where $r_{ij}$ represents
the mutual distance of body $i$ and $j$, i.e., 
%$r_{ij}=\sqrt{(\mathbf{r}_{i}-\mathbf{r}_{i})^2}$.
$r_{ij}=\sqrt{(\mathbf{r}_{i}-\mathbf{r}_{j})^2}$.
Note that
the force $\mathbf{f}_{i}$ acting on the body $i$
given by
\begin{equation}
\mathbf{f}_{i}
= -\frac{ \partial V_{\alpha}}{\partial\mathbf{r}_{i}}
=m_{i}\sum_{j \neq i} m_{j}
			(\mathbf{ r}_j-\mathbf{ r}_i)
%ii.1			r_{ji}^{(\alpha-2)}
			r_{ji}^{\alpha-2}
\label{eq:f}
\end{equation}
is a continuous function of $\alpha$
and is attractive force for all $\alpha$.
%
%repulsive force
Non-existence of motions 
with constant moment of inertia 
under repulsive forces
is obvious.
%See comment below equation \eref{eq:initialu}.
%ii.4 See comment for repulsive force in 
%ii.4 section \ref{section:summary}.
See comment for repulsive force 
in the second paragraph from the end of section \ref{section:summary}.

Without loss of generality,
we can take the centre of mass to be the origin,
\begin{equation}
\sum_{i} m_{i}\mathbf{r}_{i}(t)=\mathbf{0},
\label{eq:cm}
\end{equation}
the origin of time, t = 0, to be the instant of the condition (III),
and
\begin{equation}
\mathbf{r}_{3}(0)=\mathbf{0}.
\end{equation}
Then
the equations for the centre of mass \eref{eq:cm},
the first derivative of the moment of inertia \eref{eq:I}
%with respect to the time
with respect to time
and the zero angular momentum \eref{eq:L}
at $t=0$
yield
\begin{eqnarray}
m_{1}\mathbf{r}_{1}(0)+m_{2}\mathbf{r}_{2}(0)
	& = & \mathbf{0}, 
	\label{eq:centerOfMassAtTeq0}\\
m_{1}\mathbf{r}_{1}(0)\cdot\mathbf{v}_{1}(0)
+m_{2}\mathbf{r}_{2}(0)\cdot\mathbf{v}_{2}(0)
	& = & 0, \\
m_{1}\mathbf{r}_{1}(0)\times\mathbf{v}_{1}(0)
+m_{2}\mathbf{r}_{2}(0)\times\mathbf{v}_{2}(0)
	& = & \mathbf{0}.
\end{eqnarray}
Using the equation \eref{eq:centerOfMassAtTeq0},
let $\mathbf{a}=m_{1}\mathbf{r}_{1}(0)=-m_{2}\mathbf{r}_{2}(0)$.
Then the above equations become
\begin{eqnarray}
\mathbf{a}\cdot
	\left(\mathbf{v}_{1}(0)-\mathbf{v}_{2}(0)
	\right) & = & 0, 
	\label{eq:doteq0}\\
\mathbf{a}\times
	\left(\mathbf{v}_{1}(0)-\mathbf{v}_{2}(0)
	\right) & = & \mathbf{0} 
	\label{eq:timeseq0}
	.
\end{eqnarray}

Since
$(\mathbf{a}\cdot\mathbf{b})^2+(\mathbf{a}\times\mathbf{b})^2
=(\mathbf{a}^2)(\mathbf{b}^2)$
holds
for arbitrary vectors $\mathbf{a}$ and $\mathbf{b}$,
the equations \eref{eq:doteq0} and \eref{eq:timeseq0}
demand $\mathbf{a}=\mathbf{0}$
or $\mathbf{v}_{1}(0)=\mathbf{v}_{2}(0)$.
%trivial motion
If $\mathbf{a}=\mathbf{0}$ then $\mathbf{r}_{i}(0)=\mathbf{0}$
%for all $i=1,2,3$.
for all $i=1,2,3$ and the moment of inertia at $t=0$ is zero.
This contradicts the condition (I).
%By the condition (I),
%the moment of inertia must always be zero
%and $\mathbf{r}_{i}(t)=\mathbf{0}$.
%,
% i.e., triple collision forever.
%
%This trivial motion satisfies the conditions
%(I)--(III)
%for any attractive forces.
%
%non-trivial motion
%From now on,
%we consider non-trivial motions,
%i.e., $\mathbf{a} \neq \mathbf{0}$.
Then, we can express variables at $t=0$
as follows,
\begin{eqnarray}
\mathbf{a}=m_{1}\mathbf{r}_{1}(0)=- m_{2}\mathbf{r}_{2}(0) \neq \mathbf{0},
	\mathbf{r}_{3}(0)=\mathbf{0},\\
\mathbf{v}_{1}(0)=\mathbf{v}_{2}(0)=-\mathbf{u},
%	\mathbf{v}_{3}(0)=(m_1+m_2)m_{3}^{-1}\mathbf{u}.
	\mathbf{v}_{3}(0)=\frac{m_1+m_2}{m_{3}} \mathbf{u}.
\end{eqnarray}
Therefore,
%non-trivial 
motion under the conditions (I)--(III)
must be on a plane
defined by $\mathbf{a}$ and $\mathbf{u}$.
%ii.3
Here, it is well known that
the three body motion with
zero-angular momentum, the condition (II),
always planar \cite{Wintner, siegel}.
Using the rotation %invariance
and 
the scaling 
invariance
%property 
of this system,
we can take the Cartesian component
of these variables as follows,
\begin{eqnarray}
%\mathbf{r}_{1}(0)=(2m_2(m_1+m_2)^{-1},0),
\mathbf{r}_{1}(0)=(\frac{2m_2}{m_1+m_2},0),
	\label{eq:initialX}\\
%\mathbf{r}_{2}(0)=(-2m_1(m_1+m_2)^{-1},0),\\
\mathbf{r}_{2}(0)=(-\frac{2m_1}{m_1+m_2},0),\\
\mathbf{r}_{3}(0)=(0,0),	\\
\mathbf{u}=u(\cos\theta,\sin\theta),\ u>0,
%.
\; 0 \le \theta < 2\pi.
	\label{eq:initialV}
\end{eqnarray}
Then the kinetic 
and potential energies
at $t=0$
are given by
\begin{equation}
K(0)=\frac{(m_1+m_2)(m_1+m_2+m_3) u^2}{2m_{3}}
	\label{eq:initialK}
\end{equation}
and
%\begin{eqnarray}
\begin{equation}
\alpha V_{\alpha}(0) 
= %&&
m_{1}m_{2}2^\alpha+
	m_{2}m_{3}(\frac{2m_1}{m_1+m_2})^\alpha
	%\nonumber\\
	%&&
    +
	m_{3}m_{1}(\frac{2m_2}{m_1+m_2})^\alpha
	\mbox{ for } \alpha \neq 0.
	\label{eq:initialPotential}
%
%we don't use V_{0}(0)
%	\nonumber\\
%\fl
%=m_{1}m_{2}\log 2
%	+m_{2}m_{3}\log(2m_1/(m_1+m_2))
%	+m_{3}m_{1}\log(2m_2/(m_1+m_2))
%	\mbox{ for } \alpha = 0.	
%\end{eqnarray}
\end{equation}

%second derivative
The second derivative of the moment of inertia
with respect to time
yields the Lagrange-Jacobi identity,
\begin{eqnarray}
\label{eq:LagrangeJacobi}
\frac{d^{2}I}{dt^2}
&=& 2K-\alpha V_{\alpha}
	= 2E-(2+\alpha)V_{\alpha}
	 \mbox{ for $\alpha \neq 0$}, 
	\nonumber\\
&=& 2K-\sum_{i>j}m_{i}m_{j} 
	=2E-\sum_{i>j}m_{i}m_{j} -2V_{0}
	\mbox{ for $\alpha = 0$},
\end{eqnarray}
where $E$ represents
the total energy $E=K+V_{\alpha}$.
Thus the condition for the second derivative
$d^{2}I/dt^2=0$
yields
\begin{eqnarray}
K
&=&2^{-1}\alpha V_{\alpha}
	\ \mbox{ for } \alpha \neq 0,
	\nonumber\\
&=&2^{-1}\sum_{i>j}m_{i}m_{j}
	\ \mbox{ for } \alpha = 0.
	\label{eq:KV}
\end{eqnarray}
Note that the right-hand side of the above equation
is a continuous function of $\alpha$.
Equations
\eref{eq:initialK},
\eref{eq:initialPotential}
and \eref{eq:KV}
determine
the speed $u$ in equation
\eref{eq:initialV}
for all $\alpha$,
as follows
\begin{eqnarray}
%\fl
u^2
&=&m_{3}\Big((m_1+m_2)(m_1+m_2+m_3)\Big)^{-1}
	\nonumber\\
%\lo
&&	\times
	\Big(
	m_{1}m_{2}2^\alpha+
	m_{2}m_{3}(\frac{2m_1}{m_1+m_2})^\alpha+
	m_{3}m_{1}(\frac{2m_2}{m_1+m_2})^\alpha
	\Big).
	\label{eq:initialu}
\end{eqnarray}
%
%comment for repulsive force
%Equations for repulsive forces
%are given by
%replacing  $V_{\alpha}$ to $-V_{\alpha}$.
%Since the equation \eref{eq:LagrangeJacobi} becomes
%$d^{2}I/dt^2 = 2K+\alpha V_{\alpha} > 0$
%or
%$d^{2}I/dt^2 = 2K+\sum_{i>j}m_{i}m_{j}>0$,
%there are no motions with $I=const.$
%
%
%summary of above
As shown above, 
conditions 
$I \neq 0$, 
$dI/dt=0$, $d^{2}I/dt^2=0$
and (II)--(III)
at $t=0$
determine the initial values
with only one parameter
$\theta$ 
in equation
\eref{eq:initialV}
left undetermined.

%
%Since $d^{2}I/dt^2=2E$ for $\alpha=-2$,
%equations
%$d^{n+2}I/dt^{n+2}=0$
%do not produce any more conditions.
%%
%For the case $\alpha \neq -2$,
%on the other hand,
%%
%\[
%\frac{d^{n+2}I}{dt^{n+2}}
%=-(2+\alpha)
%	\frac{d^{n}V_{\alpha}}{dt^n}=0
%\ \mbox{ for } n=1,2,3,\cdots
%\]
%%
%give infinitely many conditions.

Higher order derivatives of $I=\mbox{const.}$,
\[
\frac{d^{n+2}I}{dt^{n+2}}
=-(2+\alpha)
	\frac{d^{n}V_{\alpha}}{dt^n}=0,
\ \mbox{ for } n=1,2,3,\cdots
\]
do not produce any more conditions for $\alpha=-2$.
On the other hand,
for the case $\alpha \neq -2$ they 
give infinitely many conditions.
%
%
%which must be satisfied by 
%non-trivial motion with (I)--(III) 
%if any
%under the potential $\alpha \neq -2$.
%
%
We call these equations at $t=0$
\begin{equation}
\frac{d^{n}V_{\alpha}}{dt^n}(0)=0
	\ \mbox{ for } n=1,2,3,\cdots
\label{eq:consistency}
\end{equation}
the consistency conditions
because these conditions must be satisfied 
by the initial values given above
if 
%non-trivial 
motion 
with conditions (I)--(III) is consistent.
%We have only one free parameter $\theta$
%for the case of equal masses,
%or three parameters $\theta$ and
%ratio between masses $m_{i}$
%for the case of general masses.

%
By virtue of the equation of motion, 
the differential operator $d/dt$ 
acting on $V_{\alpha}$
is given by
\begin{equation}
\frac{d}{dt}
=\sum_{i}\mathbf{v}_{i}\frac{\partial}{\partial\mathbf{r}_{i}}
-\sum_{i}m_{i}^{-1}\frac{\partial V_{\alpha}}{\partial\mathbf{r}_{i}}
				\frac{\partial}{\partial\mathbf{v}_{i}}.
\label{eq:dt}
\end{equation}
Using this expression
and the initial values,
we can calculate 
$d^{n}V_{\alpha}/dt^n$
at $t=0$
up to any order we want.
In the following sections
we check the consistency conditions \eref{eq:consistency},
and prove the Theorem 1 and 2.

%integration is difficult 
%but differentiation is easy
%
%you never fail with this method.
%
%if you find d^nV/dt^n \neq 0
%at some n,
%you proved the theorem.
%
%if you find d^nV/dt^n = 0
%for all n,
%you find a non-trivial solution.
%

%new section%%%%%%%%%%%%%%%%%%%%%%%%%%%%
\section{The case with three equal masses}
\label{section:equalmass}
%subsection%%%%%%%%%%%%%%%%%%%%%%%%%%%%
%\subsection{Non-conservation of Moment of Inertia 
\subsection{Inconstancy of Moment of Inertia 
with equal masses for $\alpha \ne -2, 2, 4$}
\label{section:maincase}
In this section,
we check the consistency conditions \eref{eq:consistency}
with equal masses for $\alpha \neq -2$
and prove the Theorem 1.

We take $m_{i}=1$ for all $i=1,2,3$.
Then the initial values in
\eref{eq:initialX}--\eref{eq:initialPotential}
are 
\begin{eqnarray}
\mathbf{r}_{1}(0)=(1,0),
	\mathbf{r}_{2}(0)=(-1,0),
%	\mbox{ and }
	\mathbf{r}_{3}(0)=(0,0)
	\label{eq:equalMassInitialX}
	,\\
\mathbf{v}_{1}(0)
	=\mathbf{v}_{2}(0)
	=-\mathbf{u}
	,
	\mathbf{v}_{3}(0)
	=2\mathbf{u},
	\mathbf{u}=u(\cos\theta, \sin\theta)
	\label{eq:equalMassInitialV}
,
\\
K(0)=3u^2
	,\\
V_{\alpha}(0)=
%\alpha^{-1}
%	\left(
	\frac{2^\alpha+2}{\alpha}
%	\right)
	\ \mbox{ for } \alpha \neq 0.
	\label{eq:initialPotentialEqualMass}
\end{eqnarray}
From the equation
\eref{eq:initialu},
the speed $u$ is
\begin{equation}
u=\sqrt{
	(2^\alpha+2)/6
	}
	,
	\label{eq:u}
\end{equation}
for all $\alpha$,
including $\alpha = 0$.

Let us check 
the consistency conditions \eref{eq:consistency}.
%$d^{n}V_{\alpha}/dt^{n}(0)=0$.
%
Since
the time reversal of 
the initial values
\eref{eq:equalMassInitialX}
and \eref{eq:equalMassInitialV}
is equivalent to
the $180$ degrees rotation 
of this system
around the origin
and exchange of the index 
$1 \leftrightarrow 2$
and the potential $V_{\alpha}$ is
invariant under this transformation,
the potential $V_{\alpha}$ is invariant
under the time reversal, i.e., 
$V_{\alpha}(\mathbf{r}_{i}(-t))=V_{\alpha}(\mathbf{r}_{i}(t))$.
Therefore,
all odd order derivatives
at $t=0$ vanish
\begin{equation}
\frac{d^{n}V_{\alpha}}{dt^{n}}(0)=0 \mbox{ for n=1,3,5,}\cdots.
\end{equation}

The consistency condition for the second derivative gives
\begin{equation}
0
=\frac{d^{2}V_{\alpha}}{dt^{2}}(0)
=2^{-1}
	\left(
		2+2^\alpha
	\right)
	\left(
		3(\alpha-2)\cos(2\theta)
		-(2+2^{\alpha}-3\alpha)
	\right)
\end{equation}
If $\alpha=2$, this equation is satisfied for all $\theta$.
For $\alpha \neq 2$,
this equation yields
\begin{equation}
\cos(2\theta)
=\frac{2+2^\alpha-3\alpha}{3(\alpha-2)}
=\frac{2^\alpha-2^2}{3(\alpha-2)}-1.
\end{equation}
Note that the right-hand side
is monotonically increasing continuous function
of $\alpha$,
is $1$ at $\alpha=4$,
and is larger than $-1$ for all $\alpha$.
Thus,
there is no solution of $\theta$
for $\alpha>4$, i.e.,
there are no 
%non-trivial 
motions
for $\alpha>4$.
For $\alpha \leq 4$ and $\alpha \neq 2$,
the angle $\theta$ is given by the above
equation.
Especially, 
$\cos{2\theta}=1$ for $\alpha=4$.
Thus,
initial values are completely determined
for $\alpha \leq 4$ and $\alpha \neq 2$.

%ii.6 Using these initial values,
%ii.6 we can write down 
%ii.6 the consistency conditions
%ii.6 for the fourth and sixth derivatives,
We can write down the consistency conditions for 
the fourth and sixth derivatives,
applying the derivative operator $d/dt$ given by equation \eref{eq:dt} to 
$V_{\alpha}$ four or six times 
and substituting the initial values given above as
\begin{eqnarray}
\frac{d^{4}V_{\alpha}}{dt^{4}}(0)
=\frac{(2^{\alpha}+2)f_{4}(\alpha,2^{\alpha})}{8(\alpha-2)}
=0,\\
\frac{d^{6}V_{\alpha}}{dt^{6}}(0)
=\frac{(2^{\alpha}+2)f_{6}(\alpha,2^{\alpha})}{32(\alpha-2)^2}
=0,
\end{eqnarray}
where,
\begin{eqnarray}
f_{4}(x,y)
&=&{x^2} (128-36 y+24 {y^2}+{y^3})
	-2xy(-112+62 y+5 {y^2})  \nonumber\\
&&+8 (-32-38 y+13 {y^2}+3 {y^3})
	\label{eq:deff4}\\
f_{6}(x,y)
&=&{x^4} (6144+6496y-1816 y^2+60y^3+50y^4+y^5)  \nonumber\\
&&-4x^3(10496+6520y-3676y^2+508y^3+266y^4+7y^5)  \nonumber\\
&&+4x^2(256-10288y-15032y^2+1952y^3+1846y^4+71y^5) \nonumber\\
&&-16x(-5120-10840y-9428y^2-148y^3+1186y^4+77y^5) \nonumber\\
&&+64(-448-1596y-1860y^2-299y^3+204y^4+30y^5).
	\label{eq:deff6}
\end{eqnarray}
One can easily verify that
$\alpha=2$ or $4$
are the common roots 
of $f_{4}(\alpha,2^{\alpha})=0$
and $f_{6}(\alpha,2^{\alpha})=0$.
%ii.6 In \ref{app:comroot},
%ii.6 we prove that 
%ii.6 there are no other common roots.
Moreover $f_{4}(\alpha,2^{\alpha})=0$ and 
$f_{6}(\alpha,2^{\alpha})=0$ have another root
$\alpha = -1.88...$ and $\alpha=-1.82...$, respectively.
In Appendix A,
we prove rigorously the common root of $f_{4}(\alpha,2^{\alpha})=0$
and $f_{6}(\alpha,2^{\alpha})=0$ 
are only $\alpha=2, 4$.
Therefore,
existence of 
%non-trivial 
motion 
which satisfy conditions (I)--(III)
with equal masses 
is not consistent
for $\alpha \neq -2, 2,4$.
%

%subsection%%%%%%%%%%%%%%%%%%%%%%%%%%
%\subsection{Non-trivial motions for $\alpha=2,4$}
\subsection{Motions for $\alpha=2,4$}
\label{section:alpha24}
In this section,
we give 
%non-trivial 
motions explicitly
which satisfy conditions (I)--(III)
with equal masses
under the potential $\alpha=2$ and $4$.
%ii.7
Then, we discuss the origin of these solutions
from general framework.

%alpha = 2
For $\alpha=2$,
equation \eref{eq:u} gives $u=1$.
The initial values are
\begin{eqnarray}
\mathbf{r}_{1}(0)=(1,0),
	\mathbf{r}_{2}(0)=(-1,0),
%	\mbox{ and }
	\mathbf{r}_{3}(0)=(0,0),
	\label{eq:initialXforAlpha2}\\
\mathbf{v}_{1}(0)
	=\mathbf{v}_{2}(0)
	=-\mathbf{u},
	\mathbf{v}_{3}(0)
	=2\mathbf{u},
	\mathbf{u}=u(\cos\theta,\sin\theta),
	\label{eq:initialVforAlpha2}
\end{eqnarray}
with $u=1$.
Equation of motion is
\begin{equation}
\frac{d^{2}\mathbf{r}_{i}}{dt^2}(t)
=\sum_{j \neq i}
	\left(
		\mathbf{r}_{j}(t)-\mathbf{r}_{i}(t)
	\right)
=-3 \,\mathbf{r}_{i}(t).
\end{equation}
Here, we have used $\sum_{i} \mathbf{r}_{i}(t) = 0$.
Solution is
\begin{eqnarray}
\mathbf{r}_{1}(t)
	=\left(
		\cos(\sqrt{3}\,t),
		0
	\right)
%	-2^{-1}\mathbf{r}_{3}(t),
	-\frac{1}{2}\mathbf{r}_{3}(t),
	\\
\mathbf{r}_{2}(t)
	=\left(
		-\cos(\sqrt{3}\,t),
		0
	\right)
%	-2^{-1}\mathbf{r}_{3}(t),\\
	-\frac{1}{2}\mathbf{r}_{3}(t),
	\\
\mathbf{r}_{3}(t)
	=\frac{2}{\sqrt{3}}
	\sin(\sqrt{3}\,t)
	\left(
		\cos\theta,
		\sin\theta
	\right),
\end{eqnarray}
with arbitrary angel $\theta$.
One can easily verify that
the moment of inertia is constant
$I=2^{-1} \sum_{i} \mathbf{r}_{i}^2(t)=1$.
%shape
Obviously, the shape of this motion is
not figure-eight.
This is because
figure-eight must have 
two different periods for major %axis 
and 
minor axes,
while this potential is for an isotropic harmonic oscillator.
%collision
Since 
$\mathbf{r}_{1}-\mathbf{r}_{2}
=(2\cos(\sqrt{3}\,t),0)$,
the bodies $1$ and $2$
collide at
$t=\pi/(2\sqrt{3}\,)$
for any angle $\theta$.

%alpha = 4
For $\alpha=4$,
%equation \eref{eq:u} gives $u=\sqrt{3}$.
the initial values are the same as 
equations \eref{eq:initialXforAlpha2}
and \eref{eq:initialVforAlpha2}
with
$u=\sqrt{3}$
and
\begin{equation}
	\cos(2\theta)=1.
\end{equation}
The motion is therefore
one dimensional.
%
%We take the angle $\theta$ be $0$.
We take the angle $\theta=0$.
Motion with $\theta=\pi$
is equivalent to 
time reversal motion
of $\theta=0$.
We write $\mathbf{r}_{i}(t)=(x_{i}(t),0)$
and $\mathbf{v}_{i}(t)=(v_{i}(t),0)$.
Initial conditions are
\begin{eqnarray}
x_1(0)=1, x_2(0)=-1, x_3(0)=0,\\
v_1(0)=v_2(0)=-\sqrt{3}, v_3(0)=2\sqrt{3}.
\end{eqnarray}
Equation of motion is
\begin{equation}
\frac{d^{2}x_{i}}{dt^2}(t)
=\sum_{j \neq i}
	\left(
		x_{j}(t) - x_{i}(t)
	\right)^3
=(x_j + x_k -2x_i)
	\Big(
		\sum_{\ell} x_{\ell}^2
		-
		\sum_{\ell > m} x_{\ell}x_{m}
	\Big),
\end{equation}
with $(i,j,k)=(1,2,3)$, $(2,3,1)$ or $(3,1,2)$.
Since
$\sum_{i} x_{i}(t) =0$,
this equation is reduced into 
%0.1
\cite{chen}
\begin{equation}
\frac{d^{2}x_{i}}{dt^2}(t)
%=-9/2\, x_{i}
=-\frac{9}{2}\, x_{i}
%	\times
	\Big( \sum_{j} x_{j}^2 \Big)
	.
\end{equation}
Therefore, if the moment of inertia
is constant
$I=2^{-1}\sum_{i} x_{i}^2(t)=1$,
the equation of motion is equivalent to
that of a harmonic oscillator 
\begin{equation}
\frac{d^{2}x_{i}}{dt^2}(t)=-9\, x_{i}.
\end{equation}
Solution is
\begin{eqnarray}
x_1(t)=\frac{2}{\sqrt{3}}\, \sin(3t+\frac{2\pi}{3}),\\
x_2(t)=\frac{2}{\sqrt{3}}\, \sin(3t-\frac{2\pi}{3}),\\
x_3(t)=\frac{2}{\sqrt{3}}\, \sin(3t).
\end{eqnarray}
One can easily verify that
the moment of inertia 
of this solution is constant
$I=1$,
and that
the bodies $1$ and $3$
collide at
$t=\pi/18$.
%shape
Since this motion is one dimensional,
%this motion too is not figure-eight.
this motion is not figure-eight too.

%ii.7
The origin of the above solutions are as follows \cite{chen, yoshida}.
For $\alpha=2$,
let us consider three dimensional motions with general masses $m_{i}$.
The fact that the centre of mass is at the origin
implies
\[
I = \frac{1}{2} \sum_{i} m_{i}\mathbf{ r}_{i}^2
=\frac{1}{2M} \sum_{i}m_{i}m_{j}(\mathbf{ r}_{i}-\mathbf{ r}_{j})^2
%/\sum_{i}m_{i}
=\frac{V_{2}}{M}.
\]
Here we write $M=\sum_{i}m_{i}$.
Then the Lagrange-Jacobi identity \eref{eq:LagrangeJacobi}
yields
\[
\frac{d^{2}I}{dt^2}=2E-4MI.
\]
For $\alpha=4$,
let us consider one dimensional motions with equal masses.
The situation is similar due to the identity
\[
\frac{1}{2}
	\Big(
		\sum_{i>j} (x_{i}-x_{j})^2
	\Big)^2
=\sum_{i>j} (x_{i}-x_{j})^4.
\]
Then the Lagrange-Jacobi identity yields
\[
\frac{d^{2}I}{dt^2}
=2E-27I^2.
\]
Therefore
$dI/dt=0$ and $d^{2}I/dt^2=0$
is sufficient to be $I=$constant
for both cases.
This is the reason why
higher derivatives of $I$
do not give any conditions
for initial values
as shown in the section \ref{section:maincase}.

%new section%%%%%%%%%%%%%%%%%%%%%%%%%%%%
\section{The case with general masses for $\alpha=-1$}
\label{sec:nonequalmass}
In this section,
we check
the consistency conditions \eref{eq:consistency} with general masses
under the Newtonian gravity $\alpha = -1$,
and prove the Theorem 2.

From the equation
\eref{eq:initialu},
the speed $u$ is given by
\begin{equation}
u=
\sqrt{
	\frac{
		m_3
			\left(
				m_1^2 m_2^2+
				(m_1^3+ m_1^2 m_2 +m_1m_2^2+m_2^3)m_3
			\right)
	}
	{2m_1m_2(m_1+m_2) (m_1+m_2+m_3)
	}
}.
\end{equation}
And
consistency condition for 
the first derivative gives
\begin{equation}
%\fl
0=
\frac{dV_{-1}}{dt}(0)=
-
\frac{
(m_{1}^{3}-m_{2}^{3})
%(m_1-m_2)
(m_1+m_2)^2
%(m_1^2+m_1m_2+m_2^2)
(m_1+m_2+m_3)
u
%\cos(\theta)
\cos\theta
}
{
4m_1^2m_2^2
}.
\end{equation}
This is satisfied
if $m_1=m_2$ or $\cos\theta=0$.

For the case $m_1=m_2$:
Let 
$m_1=m_2=m$ and $m_3 = \mu m$.
Then the speed $u$ becomes
\begin{equation}
u=
%2^{-1}
\frac{1}{2}
\sqrt{
\frac{m\mu(1+4\mu)}{2+\mu}
}.
\end{equation}
Since $m_{1}=m_{2}$,
the time reversal is
equivalent to
$180$ degrees rotation around the origin
and exchange the index $1 \leftrightarrow 2$.
Therefore $V_{-1}(\mathbf{r}_{i}(-t))=V_{-1}(\mathbf{r}_{i}(t))$
and the consistency conditions for odd order derivatives
are satisfied
$d^{n}V_{-1}/dt^n\,(0)=0$ for $n=1,3,5,\cdots$.
The condition for the second derivative gives
\begin{equation}
0=\frac{d^{2}V_{-1}}{dt^2}(0)=
-8^{-1}\, m^3
(1+4\mu)
\left(
	5+6\mu+6(2+\mu)\cos(2\theta)
\right).
\end{equation}
This condition is satisfied by the angle
\[
\cos2\theta = -\frac{5+6\mu}{12+6\mu}.
\]
And the condition for the fourth derivative,
\begin{equation}
0=
\frac{d^{4}V_{-1}}{dt^4}(0)=
-\frac{m^4
(1+4\mu)
(-1597 - 1576\mu + 432\mu^2)}
{384\mu},
\end{equation}
can be satisfied if
\[
\mu=\frac{197+14\sqrt{418}}{108}.
\]
But the sixth derivative is always negative,
%\begin{equation}
%\fl
%\frac{d^{6}V_{-1}}{dt^6}(0)=
%-
%\frac{
%	m^5
%	(
%	315165 + 2686088\mu + 6911872\mu^2 
%	+ 4944512\mu^3 + 443136\mu^4 +110592\mu^5
%	)
%}
%{6144\mu^2
%}.
%\end{equation}
%
\begin{eqnarray}
\frac{d^{6}V_{-1}}{dt^6}(0) &=& 
-\frac{m^5}{6144\mu^2}
	\Big(
	315165 + 2686088\mu + 6911872\mu^2 
\nonumber
\\
& &
	+ 4944512\mu^3 + 443136\mu^4 +110592\mu^5
	\Big).
\end{eqnarray}
Therefore,
$V_{-1}(t)=$const.
is not consistent in this case.

For the case
$m_1 \neq m_2$ and
$\cos\theta = 0$:
In this case,
the time reversal  is
equivalent to
%reflection about the x axis, 
reflection of the y axis, 
$y \leftrightarrow -y$.
Therefore $V_{-1}(\mathbf{r}_{i}(-t))=V_{-1}(\mathbf{r}_{i}(t))$
and the consistency conditions for odd order derivatives
are satisfied
$d^{n}V_{-1}/dt^n\,(0)=0$ for $n=1,3,5,\cdots$.
The condition for the second derivative gives
quadratic equation for $m_{3}$
\begin{equation}
0
=\frac{d^2V_{-1}}{dt^2}(0)
=-\frac{(m_1+m_2)(c_{2}m_{3}^2 - c_{1} m_{3} - c_{0})}{16m_1^3 m_2^3},
\end{equation}
with
\begin{eqnarray}
\fl
c_{2}&=&
	(m_1 - m_2)^2
	(m_1 + m_2)^2
	(m_1^2 + m_1 m_2 + m_2^2) 
	,\\
\fl
c_{1}&=&
	2m_1 m_2
	(m_1 + m_2)
	(m_1^4 + m_1^3 m_2 + 3m_1^2 m_2^2 + m_1 m_2^3 + m_2^4)
	,\\
\fl
c_{0}&=&
	m_1 m_2
	(m_1^6 + 2 m_1^5 m_2 + m_1^4 m_2^2 - m_1^3 m_2^3 
        		+m_1^2 m_2^4 + 2 m_1 m_2^5 + m_2^6)
	>0.
\end{eqnarray}
This condition is satisfied if
\begin{equation}
m_3=
\frac{
	c_1+\sqrt{c_1^2 + 4c_0 c_2}
	}{2c_2}.
\end{equation}
But the fourth derivative
has the following form
\begin{equation}
\frac{d^{4}V_{-1}}{dt^4}(0)= 
-
\frac{
3(m_1+m_2)^2 
\left(
	p_{0}(m_1,m_2)+p_{1}(m_1,m_2)\sqrt{m_1 m_2 \Omega}
	\,
\right)
}
{
q(m_1,m_2)
},
\end{equation}
where
\begin{eqnarray}
\fl
p_{0}=
7m_1^{22} + 74m_1^{21}m_2 + 
  321m_1^{20}m_2^2 + 955m_1^{19}m_2^3 + 
  2335m_1^{18}m_2^4 \nonumber\\
\lo  + 4925m_1^{17}m_2^5 + 
  9261m_1^{16}m_2^6 + 15383m_1^{15}m_2^7 + 
  22843m_1^{14}m_2^8 \nonumber\\
\lo  + 29992m_1^{13}m_2^9 + 
  35297m_1^{12}m_2^{10} + 
  37102m_1^{11}m_2^{11} \nonumber\\
\lo  + 
  35297m_1^{10}m_2^{12} + 
  29992m_1^{9}m_2^{13} + 22843m_1^{8}m_2^{14} + 
  15383m_1^{7}m_2^{15} \nonumber\\
\lo  + 9261m_1^{6}m_2^{16} + 
  4925m_1^{5}m_2^{17} + 2335m_1^{4}m_2^{18} + 
  955m_1^{3}m_2^{19} \nonumber\\
\lo  + 321m_1^{2}m_2^{20} + 
  74m_1 m_2^{21} + 7m_2^{22} ,\\
\fl
p_{1}=
21m_1^{16} + 136m_1^{15}m_2 + 
    457m_1^{14}m_2^2 + 1104m_1^{13}m_2^3  + 
    2049m_1^{12}m_2^4 \nonumber\\
\lo    + 3284m_1^{11}m_2^5 + 
    4510m_1^{10}m_2^{6} + 5516m_1^{9}m_2^{7} + 
    5830m_1^{8}m_2^{8} \nonumber\\
\lo    + 5516m_1^{7}m_2^{9} + 
    4510m_1^{6}m_2^{10} + 3284m_1^{5}m_2^{11} + 
    2049m_1^{4}m_2^{12} \nonumber\\
\lo    + 1104m_1^{3}m_2^{13} + 
    457m_1^{2}m_2^{14} + 136m_1 m_2^{15} + 
    21m_2^{16}, \\
\fl
\Omega=
m_1^{10} + 2m_1^{9}m_2 + m_1^{8}m_2^{2} + 
  4m_1^{7}m_2^{3} + 9m_1^{6}m_2^{4} + 
  15m_1^{5}m_2^{5} \nonumber\\
\lo  + 9m_1^{4}m_2^{6} + 
  4m_1^{3}m_2^{7} + m_1^{2}m_2^{8} + 
  2m_1 m_2^{9} + m_2^{10},
\end{eqnarray}
and
%
%\begin{eqnarray}
%\fl
%q=
%128
%(m_1-m_2)^4
%(m_1^2+m_1 m_2+m_2^2)^3
%(m_1 m_2)^{5/2} \nonumber\\
%\lo \times
%\left(
%	(m_1^4 + m_1^3 m_2 + 3m_1^2 m_2^2 + 
%   		m_1 m_2^3 + m_2^4
%	)
%	\sqrt{m_1 m_2}
%	+\sqrt{\Omega}
%	\,
%\right).
%\end{eqnarray}
\begin{eqnarray}
\fl
q=
128
(m_1-m_2)^4
(m_1^2+m_1 m_2+m_2^2)^3
(m_1 m_2)^{2} \nonumber\\
\lo \times
\left(
	(m_1^4 + m_1^3 m_2 + 3m_1^2 m_2^2 + 
    		m_1 m_2^3 + m_2^4
	)
	m_1 m_2
	+\sqrt{m_1 m_2\Omega}
	\,
\right).
\end{eqnarray}
Thus $d^4V_{-1}/dt^4\ (0)<0$.
Therefore,
$V_{-1}(t)=$const.
is not consistent in this case too.

%new section
\section{Summary and discussions}
\label{section:summary}
%@
In this paper, we have 
%ii.5 proved
%ii.5 ``Saari-Chenciner's Theorem":
solved
Saari-Chenciner's problem:
{\it the moment of inertia of the three-body figure-eight choreography 
under the attractive potential $V_{\alpha}$ defined in \eref{eq:V}
stays 
constant if and only if $\alpha=-2$.}
%ii.2 This ``Saari-Chenciner's Theorem" for $\alpha=-1$
This Saari-Chenciner's problem for $\alpha=-1$
is a tiny piece of the Saari's Conjecture,
because a {\it relative equilibrium} of the three bodies yields
the non-zero angular momentum
and contradicts the zero angular momentum of the figure-eight
choreography.
On the other hand, 
%ii.5 since ``Saari-Chenciner's Theorem" states the motion 
since Saari-Chenciner's problem 
states the motion 
for all $\alpha$,
% other than $\alpha=-2$,
it is considered to be a partial extension of the Saari's Conjecture.

Though the three-body figure-eight choreography 
%ii.9
with $\alpha=-2$
having constant moment of inertia
\cite{moore}%
\cite{simo1}
is not given analytically,
it will be obtained numerically from 
the set of initial conditions at $t=0$ given in
equations \eref{eq:equalMassInitialX},
\eref{eq:equalMassInitialV} and 
\eref{eq:u}, 
%ii.9 with $\alpha=-2$, 
i.e.,
\[
\mathbf{r}_{1}(0)=(1,0), \;
	\mathbf{r}_{2}(0)=(-1,0), \;
%	\mbox{ and }
	\mathbf{r}_{3}(0)=(0,0)
\]
and
\[
\mathbf{v}_{1}(0)
	=\mathbf{v}_{2}(0)
	=-\frac{1}{2}\sqrt{\frac{3}{2}} (\cos\theta, \sin\theta) 
	, \;
	\mathbf{v}_{3}(0)=\sqrt{\frac{3}{2}}(\cos\theta, \sin\theta). 
\]
Here suitable values of $\theta$ should be chosen.
%It is not known that 
The analytical method how to determine the values of $\theta$
is not known.
If a $\theta_{0}$ gives a figure-eight
then $-\theta_{0}$ and $\pi \pm \theta_{0}$
give the same figure-eight.
So we have four $\theta$ for one figure-eight.
Uniqueness of figure-eight is still unproved.

%ii.5 In order to prove the ``Saari-Chenciner's Theorem",
In order to solve the Saari-Chenciner's problem,
we have considered the motion which satisfies 
the conditions (II) and (III) instead of
the three-body figure-eight choreography.
The set of initial conditions of the motion,
$\{\mathbf{r}_{1}, \mathbf{r}_{2}, \mathbf{r}_{3},
\mathbf{v}_{1}, \mathbf{v}_{2}, \mathbf{v}_{3} \}$,
at the instant of the condition (III), $t=0$, 
has been written
by only one parameter $\theta$ in equation \eref{eq:initialV}.
The set has been determined by the conditions (II), (III), 
$I \neq 0$, $dI/dt=0$, and $d^{2}I/dt^2=0$ at $t=0$.

Since we have considered the motion
under the conditions (II) and (III), 
we have obtained the Theorem 1 as a by-product.
It is applicable to wide class of motion 
more than the figure-eight choreography.
For example, though the H3 orbit found by Sim\'{o} \cite{simo2}
is not the figure-eight choreography because each particle runs 
in the different figure-eight orbit,
it satisfies the conditions (II), (III) and $I \neq 0$.
Therefore the H3 orbit can not have constant moment of inertia
by Theorem 1.

Theorem 1 
have stated that there may exist 
%have not stated 
%non-trivial 
motions for $\alpha=2, 4$ other than $\alpha=-2$.
%ii.5 In connection with ``Saari-Chenciner's Theorem",
In connection with Saari-Chenciner's problem,
we have given the explicit solutions for $\alpha=2, 4$ and
have shown that
they are never the figure-eight choreography
because they all have collisions.

If we consider a wider class of interaction potential 
other than the power law including log potential, 
the figure-eight choreography with constant moment of inertia is possible. 
For example, under the artificial potential
$1/2 \log r - \sqrt{8}/3 r^2$,
we find a three-body choreography on the lemniscate,
a kind of analytical figure-eight.
This motion has a constant moment of inertia. \cite{fujiwara}

As we noted in section 2, 
%As we noted in section 2, below equation \eref{eq:f}, 
we comment on non-existence of the motion 
under the repulsive potential $-V_{\alpha}$
for the general mass three-body system for all $\alpha$.
This is almost obvious but can be proved 
%rigorously
since the equation \eref{eq:LagrangeJacobi} becomes
\[
\frac{d^{2}I}{dt^2} = 2K+\alpha V_{\alpha} > 0
\]
for $\alpha \ne 0$ and
\[
\frac{d^{2}I}{dt^2} = 2K+\sum_{i>j}m_{i}m_{j}>0
\]
for $\alpha = 0$ by replacing  $V_{\alpha}$ to $-V_{\alpha}$.

For the three-body system with general masses, 
the same analysis 
will be possible
but it will become more complex.
We then have done the analysis only for the 
realistic potential, the Newtonian potential $\alpha=-1$ 
and have obtained Theorem 2, 
which states that 
the motion having constant moment of inertia 
is impossible.
This is also consistent with 
the Saari's Conjecture.
Though Theorem 2 
%for the general mass three-body system for $\alpha=-1$
is still a tiny piece of the Saari's Conjecture,
it will be extended to arbitrary $\alpha$.
Actually for the equal mass system 
we could obtain the Theorem 1 and could apply it 
%ii.5 to the proof of the ``Saari-Chenciner's Theorem".
to solve the Saari-Chenciner's problem.
Extension of the Theorem 2 for all $\alpha \neq -2$ 
is left for the future work.

%appendix
\appendix
%new section
\section{Common roots of 
$d^{4}V_{\alpha}/dt^4(0)=0$ and 
$d^{6}V_{\alpha}/dt^6(0)=0$}
\label{app:comroot}
In this appendix,
we prove that
the common roots of
$f_{4}(\alpha,2^{\alpha})=0$
and $f_{6}(\alpha,2^{\alpha})=0$
are only $\alpha=2,4$.
Functions $f_{4}(x,y)$ and $f_{6}(x,y)$
are defined by equations
\eref{eq:deff4} and \eref{eq:deff6}.

By the Euclidean algorithm 
for polynomials of two variables
\cite{takagi},
we can find polynomials 
$L(x,y)$, $M(x,y)$ 
%ii.10 and $R(y)$,
and resultant $R(y)$,
which satisfy
\begin{equation}
\label{eq:euclid}
L(x,y)f_{6}(x,y)-M(x,y)f_{4}(x,y)=R(y).
\end{equation}
Actually,
\begin{eqnarray}
\fl
L(x,y)
=
x 
\Big(
	268435456+5704253440 y- 
	4900519936 {y^2}-10788732928 {y^3}  \nonumber\\
	+ 1665391872 {y^4}-2044335168 {y^5}-  
	339308448 {y^6}-16071552 {y^7} \nonumber\\
 	+3559728 {y^8}- 
	160420 {y^9}+31166 {y^{10}}+2482 {y^{11}}+31 {y^{12}}
\Big)  \nonumber\\
\lo 
-4 
\Big(
	67108864-2313158656 y-803405824 {y^2}+  
	6805321728 {y^3} \nonumber\\
	+4795789440 {y^4}-  
	1930678368 {y^5}-379246848 {y^6}-  
	11684784 {y^7} \nonumber\\
	+1953024 {y^8}+113414 {y^9}-  
	4550 {y^{10}}+2707 {y^{11}}+45 {y^{12}}
\Big),
\end{eqnarray}
\begin{eqnarray}
\fl
M(x,y)
=
{x^3} 
\Big(
	12884901888+291051143168 y+  
	129899692032 {y^2} \nonumber\\
	-865502298112 {y^3} 
	-665337083904 {y^4}+113529684992 {y^5}  \nonumber\\
	+12851181696 {y^6}-4933789824 {y^7}-  
 	907026720 {y^8}+25947264 {y^9} \nonumber\\
 	+2714152 {y^{10}}-  
	299016 {y^{11}}+79330 {y^{12}}+3288 {y^{13}}+31 {y^{14}}
\Big)  \nonumber\\
\lo 
-2 {x^2} 
\Big(
	50465865728+773060558848 y-  
	95409930240 {y^2} \nonumber\\
	-2156632178688 {y^3}- 
	752601498624 {y^4}+570103937280 {y^5}  \nonumber\\
	-53961508992 {y^6}-60918928512 {y^7}-  
	6690174624 {y^8}+59713360 {y^9}  \nonumber\\
	+43040792 {y^{10}}-1087920 {y^{11}} 
	+657030 {y^{12}}+33936 {y^{13}}+369 {y^{14}}
\Big)  \nonumber\\
\lo 
+8 x 
\Big(
	14495514624-274861129728 y-  
	257627258880 {y^2}+562210586624 {y^3}  \nonumber\\
	+1106047222784 {y^4}+651747223040 {y^5}
	-45793043520 {y^6} 
	-80250740736 {y^7} \nonumber \\
	-9184728480 {y^8}-48378720 {y^9}+67514796 {y^{10}}  \nonumber\\
	-1554720 {y^{11}}+605392 {y^{12}}+54856 {y^{13}}+715 {y^{14}}
\Big)  \nonumber\\
\lo 
-32 
\Big(
	1006632960-19713228800 y-  
	94432198656 {y^2}-4409028608 {y^3}  \nonumber\\
	+275213442304 {y^4}+281906870976 {y^5}+  
	33707210784 {y^6} \nonumber\\
	-29192050368 {y^7}  
	-4054720752 {y^8}-83636004 {y^9}+22319090 {y^{10}}  \nonumber\\
	+845634 {y^{11}}+17501 {y^{12}}+28166 {y^{13}}+450 {y^{14}}
\Big)
\end{eqnarray}
and
\begin{equation}
R(y)
=-512 (-16+y) {{(-4+y)}^4} {{(2+y)}^2}f(y),
\end{equation}
with
\begin{eqnarray}
%\fl
f(y)
&=&
-65536-10276864 y-5027392 {y^2}+  
25146656 {y^3}+27552272 {y^4} \nonumber\\
&&
 +7538528 {y^5}-  
180256 {y^6}-27646 {y^7}+944 {y^8}+21 {y^9}.
\end{eqnarray}
From the equation \eref{eq:euclid},
it is obvious that
common roots 
of $f_{4}(\alpha,2^{\alpha})=0$
and $f_{6}(\alpha,2^{\alpha})=0$
are roots of $R(y)=0$ with $y=2^{\alpha}$.

The obvious roots of $R(y)=0$ are 
$y=2^{\alpha}=4$ and $16$.
%By Sturm's Theorem
The Sturm's Theorem
\cite{takagi, Henrici} shows
%we can prove 
that $f(y)=0$ 
has only one root for $y=2^\alpha >0$.
Since
\[
f(\frac{1}{2})=-\frac{697813379}{512}<0
\]
and
\[
f(\frac{1}{\sqrt{2}})=
4286363 + \frac{66842265}{16\sqrt{2}}
>0,
\]
the positive root $y_0$ is in the interval
$1/2 < y_0 < 1/\sqrt{2}$.
Therefore
roots of $R(2^\alpha)=0$
are $\alpha=2,4$ and $\alpha_0$
with $-1 < \alpha_0 < -1/2$.

%correction 10 July --- from here ---
But we can show that
$f_{4}(\alpha,2^\alpha)=0$
does not have root
between
$-1 < \alpha_0 < -1/2$
as follows.
Let us introduce new variable $\beta=-\alpha$,
and two monotonically increasing function
$g^{(+)}(\beta)$ and $g^{(-)}(\beta)$.
\begin{eqnarray}
f_{4}(\alpha,2^{\alpha})
&=& 2^{3\alpha}
\left(
g^{(+)}(\beta)-g^{(-)}(\beta)
\right),\\
g^{(+)}(\beta)
&=&
\beta^2(
128\times 2^{3\beta}+24\times 2^{\beta}+1
)
+2\beta(
62\times 2^{\beta}+5
)\nonumber\\
&&+8(13\times 2^{\beta}+3),\\
g^{(-)}(\beta)
&=&
36\beta^2 \times 2^{2\beta}
+224\beta\times 2^{2\beta}
+16(16\times 2^{3\beta}+19\times 2^{2\beta}).
\end{eqnarray}
If $\alpha_0$ is a root of $f_4(\alpha, 2^\alpha)=0$, 
$\beta_0=-\alpha_0$ satisfies $g^{(+)}(\beta_0)=g^{(-)}(\beta_0)$
with $1/2<\beta_0 <1$.
Since
functions $g^{(\pm)}(\beta)$
are monotonically increasing functions,
we get
\begin{equation}
g^{(-)}(1/2)
<g^{(-)}(\beta_0)
=g^{(+)}(\beta_0)
< g^{(+)}(1).
\end{equation}
But actual values are
\[
g^{(-)}(1/2)=850+512\sqrt{2}>1566
\mbox{ and }
g^{(+)}(1)=1563.
\]
This is a contradiction.
% --- to here ---

Thus,
we proved that
common roots of
$f_{4}(\alpha,2^\alpha)=0$
and 
$f_{6}(\alpha,2^\alpha)=0$
are only
$\alpha = 2, 4$.

\ack
We thank
Suehiro Kato
for valuable discussions.

\end{document}